\documentclass[11pt]{article}
\input amssym.def
\input amssym.tex
\usepackage{epsfig}

\def\be{\begin{equation}}
\def\ee{\end{equation}}
\def\ba{\begin{eqnarray}}
\def\ea{\end{eqnarray}}

\begin{document}
\bigskip
\title{{\bf Higher Dimensional Taub-NUTs and Taub-Bolts in Einstein-Maxwell Gravity}}

    \author{ \large Adel M. Awad\thanks{email: aawad@ictp.it } \,\,$^{1,2}$ \\
    \\
    $^{1}$ Abdus Salam International Centre for Theoretical Physics, \\ Strada Costiera 11, 34100, Trieste, Italy. \\
    \\
    $^{2}$ Department of Physics, Faculty of Science, Ain Shams University, \\Abbassia, Cairo 11566,
    Egypt.}

    \maketitle

\begin{abstract}
We present a class of higher dimensional solutions to
Einstein-Maxwell equations in $d$--dimensions. These solutions are
asymptotically locally flat, de--Sitter, or anti--de Sitter
space-times. The solutions we obtained depend on two extra
parameters other than the mass and the nut charge. These two
parameters are the electric charge $q$ and the electric potential
at infinity, $V$, which has a non-trivial contribution. We Analyze
the conditions one can impose to obtain Taub-Nut or Taub-Bolt
space-times, including the four-dimensional case. We found that in
the nut case these conditions coincide with that coming from the
regularity of the one-form potential at the horizon. Furthermore,
the mass parameter for the higher dimensional solutions depends on
the nut charge and the electric charge or the potential at
infinity.

\end{abstract}

\section{Introduction}

The importance of Taub-Nut solutions covers a wide area of
application that extends from general relativity to string theory.
These solutions \cite{original} are characterized by non-vanishing
nut charges. As a result, they are locally asymptotically flat,
i.e., their boundaries are not $S^1\times S^2$ but $S^1$ fiber
over $S^2$. These boundaries are topologically interesting since
they do have a non-vanishing first Chern number $N$ which is
proportional to the nut charge. In these space-times one can not
define a global time function, therefore, these manifolds can not
be foliated using constant time surfaces and it is not possible to
describe their time evolution through a unitary Hamiltonian
evolution. Taub-Nut spaces are also characterized by the existence
of zero-dimensional fixed point set of the $U(1)$ isometry
generated by the time-like Killing vector $\partial_{\tau}$, which
is called a nut. This is in contrast with the higher-dimensional
fixed-point set appears in spherically symmetric black hole
solutions, which is called a bolt. Tau-Nut solutions also possess
a singular one dimensional string analogous to Dirac string
\cite{misner}which is called Misner string. The singularity of
Misner string is a coordinate singularity, since one fails to
describe the manifold using a single coordinate system. Misner
string contributes to the entropy of the black hole, therefore the
entropy is not simply one quarter the area of the horizon. This is
a very interesting property in gravitation, which reveals some
features of the geometric entropy and has been studied by Hawking
and Hunter \cite{hunter,hawking+hunter,page+hunter+hawking}.

Another attractive feature of the Taub-Nut solution is the
celebrated Gross-Perry-Sorkin monopole\cite{GPS}. This monopole
can be constructed by adding another trivial dimension to get a
five dimensional metric followed by Kaluza Klein compactifing the
theory to four-dimensional along the Euclidian time direction. As
a result, one gets a singular four-dimensional monopole solution.
This shows how perfectly regular higher dimensional solutions can
give rise to a singular lower dimensional one.

Taub-Nut solution plays an important role in revealing the nature
of D6-branes in type IIA string theory and its relation to
M-theory(for a review see {\it e.g.}\cite{Clifford-D-book}).
Probing the background geometry of N coincident D6-branes using
D2-branes shows that the transverse dimensions to both branes are
not three, but in fact four dimensions and the metric is the
Taub-Nut metric . The extra dimension is the Euclidian time
direction which is compact. It comes from the gauge degrees of
freedom of the $U(1)$ gauge theory living on the D2-brane
world-volume. This shows that the D6-brane is a Kaluza-Klein
monopole from the M-theory prospective. Another interesting
application of Nut charged space-times is to extend and test the
validity of the AdS/CFT correspondence to locally asymptotically
AdS solutions, and study their
thermodynamics\cite{CRRC,adel+andrew,mann}.
%%From the field theory point of veiw ****This fits very well with the fact that the
%%moduli space of the $2+1$ gauge theory that live on the world-volume of the D2-brane with eight supercharges is
%%hyperkhaller

Therefore, it is interesting to find new Taub-Nut and Taub- bolt
solutions in higher dimensions with an eye on possible
applications of these results in various supergravity/string
theories. Here we present a class of higher dimensional $U(1)$
electrically charged solutions with nut charge, which are
generalizations to the solutions found in \cite{bais,adel+andrew}
(see also \cite{mann-thermo} for discussions on their
thermodynamics).

The action for Einstein-Maxwell theory in $d$--dimensions for
asymptotically (anti)--de--Sitter space-times is

\begin{eqnarray}
I_d= -{1 \over 16 \pi G_d}\int_{\cal M} d^{d}x
\sqrt{-g}\left(R+{(d-1)(d-2) \over
l^2}-F^{\mu\nu}F_{\mu\nu}\right),
\end{eqnarray}
where $\Lambda{=}{-(d-1)(d-2)\over 2l^2}$ is the cosmological
constant, $G_{d}$ is Newton's constant in $d$--dimensions, and
$F_{\mu\nu}=\partial_{\mu}A_{\nu}-\partial_{\nu}A_{\mu}$, where
$A_{\mu}$ is the vector potential. Varying the action with respect
to the metric and the vector field $A_{\mu}$ we get the following
field equations \ba
G_{\mu\nu}-\Lambda g_{\mu\nu}&=&2\,T_{\mu\nu},\nonumber\\
\partial_{\mu}\left(\sqrt{-g} F^{\mu\nu}\right)&=&0,
\ea where \be T_{\mu\nu}=F_{\mu \rho}\,F_{\nu}^{\rho}-{1 \over
4}g_{\mu\nu}F^{\mu\nu}F_{\mu\nu}. \ee

Here we consider only the Euclidian sections of these metrics. To
go to the Lorentzian sections one can analytically continue the
coordinate $\tau$ and the parameter $n$ (i.e., $\tau \rightarrow
i\,t$, $q \rightarrow i\,q$, $V \rightarrow i\,V$ and $n
\rightarrow i\,n$). The general form of the Taub-Nut/Bolt metric
is given by \be ds^2=F(r)^{-1}dr^2+(r^2-n^2)\,d\Sigma_{\cal
B}^2+F(r)(d\tau+{\cal A})^2\ee where the metric $d\Sigma_{\cal B
}^2$ is over an even dimensional Einstein-K\"{a}hler manifold
${\cal B}$, $F(r)$ is some function of $r$. $n$ is called the "Nut
charge" and ${\cal A}$ is the potential of the K\"{a}hler form
${\cal F}$ \be {\cal F}=d {\cal A} \ee

The hypersurface defined by $r=constant$ is a $U(1)$ fiber over
the base manifold ${\cal B}$. The base space ${\cal B}$ can be any
even dimensional Einstein-K\"{a}hler manifold. The case when
${\cal B}={\Bbb C}{\Bbb P}^n$ is special, since the solution
obtained is a non-singular space. Any other base space will be
singular and therefore it can not be considered as a gravitational
instanton. On the other hand these solutions are fine when they
describe Bolt solutions, since these singularities will be hidden
behind horizons. In the ${\Bbb C}{\Bbb P}^n$ case the resulting
hypersurfaces are squashed $S^{d-1}$ and $n$ here measures the
amount of squashing.

In order to generalize this construction by giving electric
charges to these solutions, we need some ansatz for the one-form
potential $A$. We are going to adopt the following ansatz for the
potential
 \be A=g(r)\,(d\tau+{\cal A}) \ee The general solution for $g(r)$
depends on two parameters, one is the electric charge, $q$ and the
other is $V$, which can be viewed as potential at infinity; As
$r\rightarrow \infty$ we have \be A=(q/r^{d-3}+V)\,(d\tau+{\cal
A}). \ee $V$ plays an important role in studying the
thermodynamics of Riessner-Nordstrom (A)dS solutions as has been
revealed in \cite{charge-cliff}. Requiring regularity of the
one-form potential $A$ at the horizon relates the two parameters
$V$ and $q$. In fact we will not discuss the de--Sitter versions
of these solutions, but they can be obtained, simply by taking
$l\rightarrow i\,l$ which will be left for future work. While
writing this article the work of \cite{new} has been posted. The
authors of \cite{new} discuss similar solutions to the one
presented here, but one can spot two general differences; i) the
solutions presented here are a bit more general since it has two
extra parameters $V$ and $q$, ii) here we also discussed the
conditions of having Taub-Nut or Taub-Bolt solutions and we have
showed the possibility of having a Taub-Nut for certain values of
the mass parameter and $V$. This article is organized as follows;
in the first section we discuss conditions for obtaining nut and
bolt solutions for asymptotically locally flat or locally AdS
space-times in four dimensions. In the following section we
present our results in six dimensions which can be generalized to
eight and ten dimensions for asymptotically flat or AdS
space-times. Finally we put some final remarks on the relevance of
these solutions.

\section{Four-Dimensional Solutions}

\subsection{Charged Taub-Nut-(A)dS Solution}

The first electrically charged four-dimensional Taub-Nut solution
has been introduced in \cite{brill}. More general versions of this
solution have been studied in e.g., \cite{tnkerrnew,clif+myers}.
Their supersymmetry has been discussed in \cite{tnkerrnew}, where
the authors have showed that these solutions do preserve some
supersymmetry for certain choices of their parameters. Here we
present an (A)dS version of this solution with the additional
parameter, V, that we mentioned in the introduction\footnote{As it
will be clear from the discussion below this solution is
equivalent to some of the solutions discussed in
\cite{tnkerrnew}}. The charged Taub-Nut (Anti)-de-Sitter space
version of the Brill's solution has the form \be
ds^2=f(r)\,(d\tau-2n\cos{\theta}\,d\phi)^2+{dr^2 \over f(r)}+(r^2
-n^2)(d\theta^2+sin{\theta}^2\,d\phi^2)\ee where $f(r)$ is given
by \be
f(r)={r^4+(l^2-6n^2)\,r^2-2m\,l^2\,r-l^2\,(q^2+4q\,V\,n\,k+4\,n^2\,V^2\,(k^2-1)-n^2)-3n^4 \over l^2(r^2-n^2)}.\\
\ee The gauge potential has the form \be
A=g(r)\,(d\tau+2n\,\cos{\theta}d\phi). \ee Here $g(r)$ is given by
\be g(r)=-V\,{r^2+n^2+2\,n\,k\,r \over r^2-n^2}-q\,{r \over
r^2-n^2},\\\ee $q$ is the electric charge, $m$ is the mass
parameter, $k$ is some constant\footnote{In order to have a real
one-form $A$ and $f(r)$ upon going to the Lorentzian section one
must set $k=0$} and $V$ is the potential at infinity.

In order for this solution to describe a nut, we must have
$f(r=n)=0$, so that all of the extra dimensions collapse to zero
size at the fixed-point set of the Killing vector,
$\partial_{\tau}$. This only happens at a specific value of the
mass parameter, $m$, and the parameter $q$, namely\be
m_n=n-4n^3/l^2 \ee \be q = -2n\,V\,(1+k) \ee otherwise the
solution will have a horizon radius $r_h> n$ and become a bolt.
Amusingly the last condition, i.e., $q = -2n\,V\,(1+k)$ can be
obtained from requiring the regularity of the one-form $A$ at the
horizon as well. In fact this is might be a sign of consistency
for including the extra parameter, $V$. Satisfying the above
conditions gives $f(r)$ the following form \be f(r)={(r-n)\over
(r+n)l^2}\,[(r-n)\,(r+3n)+l^2)].\ee Notice here that the roots of
$[(r-n)\,(r+3n)+l^2)]$ are all less than $n$.

Since the fiber has to close smoothly at $r=n$, we do that by
setting the period $\beta$ of the $\tau$ direction to be
$\beta={4\pi\over f'(r=n)}$. This implies that \be \beta
=8\pi\,n\ee

Notice that the value of the mass parameter that produces the nut
in the charged case is identical to that of the uncharged case.
Furthermore the stress-energy tensor vanishes upon satisfying the
above conditions. As a result the metrics for the two cases in
four-dimensions are the identical. It will be clear from the
discussion below that this is not the case for higher dimensional
solutions. In higher dimensions the mass parameter of the charged
case will depend on both the nut charge and the electric charge.
This is a new feature that makes higher dimensional charged nut
solutions more richer and possess one more parameter, namely
$q$/or $V$, as it is clear from curvature tensor calculation.

Another feature which is unique to the four-dimensional solution
is that the existence of the additional parameter $V$ together
with the nut charge induces a magnetic charge and changes electric
charge as well even at large radius.  As we go to the asymptotical
flat limit $n\rightarrow 0$ the solution will have no magnetic
charge. The magnetic monopole appears as a result of the
non-trivial fibration. Also in this limit the electric charge will
be only $q$. This shows that the parameters $q$ and $V$ are not
convenient for the four-dimensional solution and they should be
replaced by $Q$ and $P$, the total electric charge and the
magnetic charge of the solution as in \cite{tnkerrnew}, which
shows that this solution is a dyon. This phenomenon does not
happen in higher dimensions, i.e. the $q$ will be the total
electric charge up to some numerical factor without any
contribution from $V$, but it is clear from the electric field
calculation that there is a linear charge density term as well
which is proportional to $n^2\, V$. As a result we keep
parameterizing these solutions using $q$ and $V$ in the higher
dimensional solutions.

\subsection{Charged Taub-Bolt-(A)dS Solution}
To have a regular bolt solution at the horizon, i.e., at
$r=r_b>n$, the following two conditions must be satisfied
simultaneously: \\
(a) $F(r_b)=0$\\
(b) $F'(r_b)={1 \over 2\,n}$\\
The first condition is just the definition of a horizon and the
second is following from the fact that we need to avoid conical
singularities at the bolt, keeping at the same time the
periodicity of $\tau$ to be $8\, \pi \,n$. In addition to these
requirements we have another regularity requirement, namely;
regularity of the one-form potential $A$ at the bolt $r=r_b$. This
leads to the following relation \be q =
-V\,{({r_b}^2+n^2+2\,n\,{r_b}\,k)\over {r_b}}\ee Imposing
condition (a), the mass parameter is given by \be
m=m_b=1/2\,{[{r_b}^4+(l^2-6\,n^2)\,{r_b}^2+4\,l^2\,n^2\,V^2\,(1-k^2)-3\,n^4+l^2\,n^2-4\,q\,k\,l^2\,V\,n-q^2\,l^2]\over
l^2\,{r_b}} \ee By imposing condition (b) one gets $r_b$ as the
solution of the following fourth order algebraic equation,

\be
6\,{r_b}^4\,n-{r_b}^3\,l^2+(2\,V^2\,l^2\,n+2\,n\,l^2-6\,n^3)\,{r_b}^2-2\,n^3\,V^2\,l^2=0\ee

One can choose to write the previous expressions in terms of
either $q$ or $V$. Here we have chosen $V$ since it is a bit
easier, otherwise the order of the algebraic equation will be
higher. One can check numerically that there are solutions (with
$r_b > n$)to the above algebraic equation in certain ranges of $V$
and $n$ for a fixed $l$. But unless we get some expression for the
Entropy and the Specific heat we can not estimate the correct
ranges of the above parameters and we leave that for a future
work.

\subsection{Charged Taub-Nut and Taub-Bolt Solution}

Here we are presenting the locally asymptotically flat version of
the above solutions. The four dimensional solution has the form
\be ds^2=f(r)\,(d\tau-2n\cos{\theta}\,d\phi)^2+{dr^2 \over
f(r)}+(r^2 -n^2)(d\theta^2+sin{\theta}^2\,d\phi^2)\ee where $f(r)$
is given by \be
f(r)={r^2-2m\,r-(q^2+4q\,V\,n\,k+4\,n^2\,V^2\,(k^2-1)-n^2)\over (r^2-n^2)}.\\
\ee The gauge potential has the form \be
A=g(r)\,(dt+2n\,\cos{\theta}d\phi). \ee Here $g(r)$ is given by
\be g(r)=V\,{r^2+n^2+2\,n\,k\,r \over r^2-n^2}+{q\,r \over
r^2-n^2},\\\ee.

As before, in order for this solution to describe a nut solution,
we must have $f(r=n)=0$, so that all of the extra dimensions
collapse to zero size. This only happens at specific value of the
mass parameter, namely \be m_n=n\ee otherwise the solution will
have a horizon radius $r_h > n$ and will describe a bolt. Choosing
the mass parameter to have the above value, $f(r)$ gets the
following form \be f(r)={(r-n)\,(r+n)}\ee

In order to get the fiber to close smoothly at $r=n$, one should
set the period $\beta$ of the $\tau$ direction to be
$\beta=4\pi\over f'(r=n)$. This implies that \be \beta
=8\pi\,n.\ee As we have mentioned the existence of the parameter
$V$ together with the nut charge induce a magnetic charge $q
\propto V\,n$ which vanishes in the asymptotic flat limit
$n\rightarrow 0$.

\begin{figure}
\begin{center}
\epsfig{file = 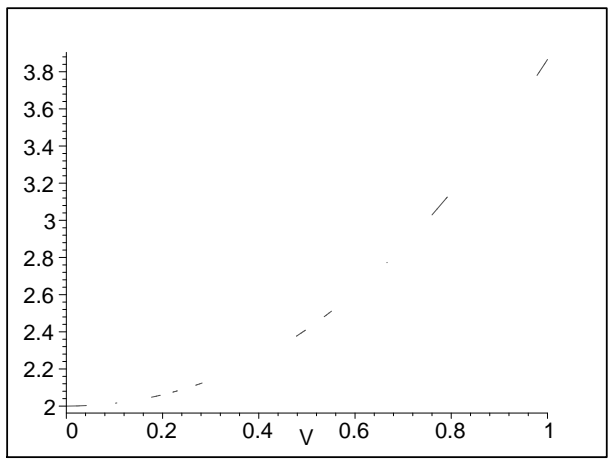, height=6cm,angle=0,trim=0 0 0 0}\caption{
$r_b$ as a function of V, $n=1$.} \label{rootplot}
\end{center}
\end{figure}

In the case of bolt solution, we impose the regularity condition
that the one-form potential $A$ vanishes at the bolt $r=r_b$. It
leads to the same relation \be q =
-V\,{({r_b}^2+n^2+2\,n\,{r_b}\,k)\over {r_b}}\ee Also, imposing
condition (a), the mass parameter is given by \be
m=m_b=1/2\,{[{r_b}^2+4\,n^2\,V^2\,(1-k^2)+n^2-4\,q\,k\,V\,n-q^2]\over
{r_b}} \ee By imposing condition (b) one gets $r_b$ as the
solution of the following fourth order algebraic equation,

\be -{r_b}^3+2\,n\,(\,V^2+1\,)\,{r_b}^2-2\,n^3\,V^2=0\ee In fact
one can solve the above equation analytically but instead of
presenting the relatively complicated expression we plotted $r_b$
in \ref{rootplot} as a function of $V$, where $n=1$. One can see
clearly that $r_b \geq 2n$.

\section{Six-Dimensional Solutions}
There are two different choices for the base manifold in six
dimensions, namely; $ {\Bbb C}{\Bbb P}^2$ and $S^2\times S^2$. Let
us start with the $ {\Bbb C}{\Bbb P}^2$ base manifold.
\subsection{Charged Taub-nuts-AdS with ${\cal B} = {\Bbb C}{\Bbb P}^2$}
Using $ {\Bbb C}{\Bbb P}^2$ as a base space, the metric of the
charged Taub-NUT-AdS solution has the form

\begin{eqnarray}
    ds^2 &=& f(r)(d\tau+{u^2 \over 2(1+u^2/6)}(d\psi+\cos\theta d\phi))^2+
    f(r)^{-1}dr^2+(r^2-n^2)d{\Sigma_2}^2,
    \end{eqnarray}
    where $d{\Sigma_2}^2$ is the metric over ${\Bbb C}{\Bbb P}^{2}$
    which has the following form
    \begin{eqnarray}
    d{\Sigma_2}^2&=&{du^2 \over (1+u^2/6)^2}+{u^2 \over
    4(1+u^2/6)^2}(d\psi+\cos\theta d\phi)^2+{u^2 \over
    4(1+u^2/6)}(d\theta^2+\sin^2\theta d\phi^2),
    \end{eqnarray}

where f(r) is given by \ba f(r)=
&&\frac{1}{6\,l^2\,(r^2-n^2)^4}\,\left[6\,r^{10}+(2\,l^2-42\,n^2)\,r^8+(156\,n^4-16\,n^2\,l^2-24\,n^2\,l^2\,V^2)\,r^6\right.\nonumber\\
&&+(20\,n^4\,l^2-180\,n^6-360\,n^4\,l^2\,V^2)\,r^4-96\,r^3\,n^2\,l^2\,q\,V+(30\,n^8-9\,l^2\,q^2+216\,n^6\,l^2\,V^2)\,r^2\nonumber\\
&&\left.+3\,n^2\,l^2\,q^2-6\,n^8\,l^2+30\,n^{10}-216\,n^8\,l^2\,V^2\right]-\frac{2\,m\,r}{(r^2-n^2)^2}\ea
The gauge potential has the form

\be A=g(r)(d\tau+{u^2 \over 2(1+u^2/6)}(d\psi+\cos\theta
d\phi))\ee

where $g(r)$ is given by \be g(r)= q \, \frac{
r}{(r^2-n^2)^2}+V\,\frac{-r^4+6\,r^2\,n^2+3\,n^4}{(r^2-n^2)^2}\ee

This solution is a generalization to the solutions found in
\cite{adel+andrew,page+pope}. The thermodynamics for the uncharged
version of these solutions have been studied in \cite{mann}.

For this solution to describe a nut, one must impose $f(r=n)=0$,
this happens when the mass parameter takes the value \be m_n=
\frac{4\,n^3\,(6\,n^2-l^2-9\,l^2\,V^2)}{3\,l^2}\ee otherwise the
solution will have a horizon radius $r_h > n$ and the solution
will describe a bolt. Notice here that the value of the mass that
causes the nut is not the same as the one for the uncharged
solution, as we mentioned in the previous section, since we have a
dependence on either $V$ or $q$. Choosing the mass parameter to
have the above value $f(r)$ will have the following form \ba
f(r)=&&\frac{(r-n)}{3\,(n+r)^4\,l^2}\left[3\,r^5+15\,n\,r^4+(24\,n^2+l^2)\,r^3+5\,n\,l^2\,r^2\right.\nonumber\\
&&-\left.(27\,n^4+12\,n^2\,l^2\,V^2-7\,n^2\,l^2)\,r-15\,n^5+3\,l^2\,n^3+12\,n^3\,l^2\,V^2
\right]\ea The regularity of the potential $A$ at $r=n$ require
that \be q= -8\,n^3\,V.\ee Again, here one can choose to write the
previous expressions in terms of either $q$ or $V$. As before we
are going to write expressions in terms of $V$.

The fiber has to close smoothly at $r=n$, this can be achieved by
setting the period $\beta$ to be \be \beta= 12 \pi \, n.\ee In
order not to get any solution with $r_b\neq \, n$ which might
spoil the nut solution we must demand that the expression in the
square brackets in Eqn. (3.38) does not have any positive roots
greater than $n$. This can be achieved by setting $r=n+x$ in this
expression which assumes the form \ba
&&[3\,x^5+30\,n\,x^4+(l^2+114\,n^2)\,x^3+(192\,n^3+8\,n\,l^2)\,x^2\nonumber\\
&&+(-12\,n^2\,l^2\,V^2+120\,n^4+20\,n^2\,l^2)\,x+16\,l^2\,n^3],
\ea then by requiring $|V| < 1/3\,{\sqrt{90\,n^2+15\,l^2}/
l}$\footnote{This is a more stronger condition but it shows most
of the range at which $V$ can produce a nut} in this expression,
it will have no positive roots. The above requirement might serve
as a condition on $V$ for obtaining a nut solution in this case.
There is no four dimension analog for this condition. As we have
mentioned in the introduction for ${\cal B} = {\Bbb C}{\Bbb P}^2$
case Taub-Nut solution have no curvature singularities at $r=n$,
indeed, \be R^{ijkl}R_{ijkl} ~{\sim}~ (r+n)^{-12}\ee

\subsection{Charged Taub-Bolt-AdS $ {\Bbb C}{\Bbb P}^2$ Case}
As in Four dimensions, in order to have a regular bolt solution at
the horizon, i.e., at $r=r_b>n$, the following two conditions must
be satisfied
simultaneously: \\
(a) $F(r_b)=0$\\
(b) $F'(r_b)={1 \over 3\,n}$\\
The first condition is just the definition of a horizon and the
second is following from the fact that we need to avoid conical
singularities at the bolt, keeping at the same time the
periodicity of $\tau$ to be $12\, \pi \,n$. In addition to these
requirements we have another regularity requirement, namely;
regularity of the one-form potential $A$ at the bolt $r=r_b$. This
leads to the following relation \be V=-{q\,r_b \over (r_b-n)^2}\ee
Imposing condition (a), the mass parameter is given by \be
m=m_b=-1/2\,{(-{r_b}^4+6\,n^2\,{r_b}^2-l^2\,{r_b}^2+3\,n^4-l^2\,n^2+4\,q\,l^2\,V\,n+q^2\,l^2)\over
(l^2\,{r_b})} \ee By imposing condition (b) together with the
regularity of the one-form $A$ one gets $r_b$ as the solution of
the following fourth order algebraic equation \be
30\,n\,{r_b}^4-2\,{r_b}^3\,l^2+(6\,n\,l^2+27\,n\,l^2\,V^2-30\,n^3)\,{r_b}^2-27\,n^3\,l^2\,V^2=0\ee
\subsection{Charged Taub-Nut and Bolt for the $ {\Bbb C}{\Bbb P}^2$ Case}
Taking the limit $l\rightarrow \infty$ one can obtain the
asymptotically locally flat version of the above solutions which
represents also new solutions of Einstein-Maxwell equations in six
dimensions. Here $f(r)$ takes the following form \ba F(r)=
&&\frac{1}{6\,(r^2-n^2)^4}\,\left[2\,r^8-(16\,n^2+24\,n^2\,V^2)\,r^6\right.\nonumber\\
&&+(20\,n^4-360\,n^4\,V^2)\,r^4-96\,r^3\,n^2\,q\,V-(9\,q^2-216\,n^6\,V^2)\,r^2\nonumber\\
&&\left.+3\,n^2\,q^2-6\,n^8-216\,n^8\,V^2\right]-\frac{2\,m\,r}{(r^2-n^2)^2}\ea
The one-form potential $A$ and the function $f(r)$ are the same as
the AdS case. For the nut case the mass parameter, $m$, and the
charge, $q$, must acquire the following form \be m_n=-{4\,
n^3\over 3}\ee \be q=-8\,n^3\,V\ee Imposing the regularity
condition for $A$ and shift $r=x+n$, one gets the condition for
getting a nut solution \be |V| < \sqrt{5\over 3}\ee

\begin{figure}
\begin{center}
\epsfig{file = 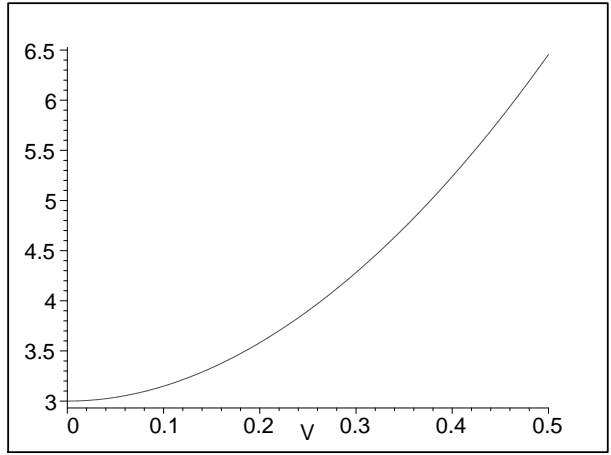, height=6cm,angle=0,trim=0 0 0
0}\caption{ $r_b$ as a function of V, $n=1$.} \label{root2plot}
\end{center}
\end{figure}

For the Taub-Bolt case imposing the general conditions a) and b)
together with the regularity condition on the potential $A$ the
horizon radius satisfy the following algebraic equation \be
2\,{r_b}^3-3\,n\,(2+9\,V^2)\,{r_b}^2-27\,n^3\,V^2=0\ee Again the
analytic solution of this equation is a relatively complicated
expression so, instead, we plot $r_b$ in \ref{root2plot} as a
function of $V$, where $n=1$. One can see clearly that $r_b \geq
3n$.

\subsection{Solutions with ${\cal B} = S^2\times S^2$}
Using $S^2\times S^2$ as a base space, the metric of the charged
Taub-NUT-AdS solution has the form

\be ds^2= f(r)(d\tau+
2n\cos{\theta_1}d\phi_1+2n\cos{\theta_2}d\phi_2)^2
+f(r)^{-1}dr^2+(r^2-n^2)({d\theta_1}^2+\sin{\theta_1}^2d\phi_1^2+{d\theta_2}^2+\sin{\theta_2}^2d\phi_2^2)\ee

\be
A=g(r)(d\tau+2\,n\,\cos{\theta_1}d\phi_1+2\,n\,\cos{\theta_2}d\phi_2)\ee
where $g(r)$ is given by \be g(r)= q \, \frac{
r}{(r^2-n^2)^2}+V\,\frac{-r^4+6\,r^2\,n^2+3\,n^4}{(r^2-n^2)^2}\ee

Here $f(r)$ has the same form as the one for $ {\cal B}={\Bbb
C}{\Bbb P}^2$ case. This Six dimensional solution is a
generalization to the solution found in
\cite{adel+andrew,page+pope}. Another exotic form similar to these
metrics has been found in \cite{nutty}. Also, these uncharged
solutions have been used to find some bubble solutions in
\cite{bubbleman}. For ${\cal B} = S^2\times S^2$ case Taub-Nut
solution have curvature singularities at $r=n$, indeed, \be
R^{ijkl}R_{ijkl} ~{\sim}~ (r-n)^{-2}\ee But the Taub-Bolt cases
for such solution do not suffer from such a problem since the
singularity is hidden behind the horizon.

\section{Final Remarks}

We have presented a class of $d$--dimensions nut-charged solutions
for Einstein-Maxwell field equations. These solutions depend on
two extra parameters other than the mass and the nut charge,
namely; the charge $q$ and the potential at infinity $V$. The
existence of the parameter $V$ enables us to get a regularity
condition on the one-form potential which we found identical to
that required to obtain a nut solution. The nut conditions show
that the mass not only depends on the nut charge as in four
dimensions but also on the charge/or the potential $V$ as a result
we get a smooth six dimensional manifold for ${\cal B}={\Bbb
C}{\Bbb P}^2$. One might be concerned by the existence of naked
singularities for the ${\cal B} = S^2\times S^2$ case. However,
one would argue that from the point of view of the adS/CFT duality
that it is possible to resolve these singularities, for example,
using Gubser's \cite{steve} criterion that these are `good'
singularities, in the sense that a finite temperature deformation
will yield a non-singular solution which is the bolt. It will be
interesting to probe the physics of these new smooth gravitational
solutions especially in the context of AdS/CFT correspondence and
their thermodynamics. One can generalize these constructions
easily to eight and ten dimensions. It is interesting also to
generalization this construction to include black holes with
topological horizons and study the nut and bolt conditions.
\\

{\noindent \bf \large Acknowledgements}\\

I would like to the thank the Abdus Salam ICTP for its hospitality
and continues support, this is where this work was accomplished. I
would like to thank Edward Teo and  Nidal Chamoun for useful
discussions, especially NC for suggestions and comments on a
preliminary draft of this paper.

\end{document}